# Features of amplification of dipole magnetic field with linear ferromagnetic concentrator

S. I. Bondarenko,[a)] A. A. Shablo, and V. P. Koverya
*Institute for Low Temperature Physics and Engineering, NASU, 47 Lenin Avenue 61103, Kharkov, Ukraine*

Yusheng He, Fenghui Zhang, and Hongsheng Ding
*Institute of Physics, Chinese Academy of Sciences, 8 Zhong Guan Cun Nan San Jie, P.O. Box, 603, 100080 Beijing, China*

Amplification of a low frequency magnetic field from a dipole source by means of linear ferromagnetic concentrator of different dimensions (permalloy micro wire) has been investigated experimentally. The results obtained strengthen the validity of the analytical relation for amplification which allows for nonlinear dependence of the magnetic permeability upon the value and distribution of the dipole field over concentrator length.

## I. INTRODUCTION

The sensitivity of magnetic measurements has greatly improved during the recent decades, which has stimulated the advent of superconducting quantum interference devices (SQUIDs), high-sensitivity Hall probes, ferromagnetic tunnel junctions, and so on. Linear ferromagnetic concentrator[1] (LFC) offers another possibility of upgrading the present-day magnetic detectors by increasing their sensitivity to homogeneous fields. Of equal importance is the sensitivity to inhomogeneous magnetic field, first of all to the field of dipole sources. This problem is a real challenge in making nonforce magnetic microscopes. The Ukrainian coauthors of this article were the first (1995) to offer LFC as a means of improving the sensitivity of SQUID-based magnetic microscopes.[2,3] The arrangement includes a dipole source with magnetic moment $M_G$ at the surface of the examination object, LFC, and a SQUID-based detector. The use of LFC in [conjugation with a SQUID-based detector provides the maximum amplification and rules out extra LFC-induced magnetic noise in the region near the SQUID detector. Subsequent investigations by other researchers[4,5] lent more evidence in favor of this approach.

The goal of our experiments was to investigate closely how the LFC parameters and the $M_G$ value influence the field amplification at the LFC end situated in the region near the SQUID detector. An attempt was also made to derive on the basis of experimental data at least an approximate analytical relation that can be used to calculate amplification.

## II. EXPERIMENTAL TECHNIQUE

Permalloy microwires (79%Ni) of two diameters (0.1 and 0.3 mm) with the highest permeability $\mu_{m,\max}=500$ at the frequency of magnetization reversal $f=10^4$ Hz were used as LFC samples.

The dipole was formed using a miniature copper ac coil (I.D.=0.1 mm or 0.3 mm). The coil was wound on one of the LFC ends (generating coil, $G$). The frequency of the alternating current $I$ in the coil was $f=10^3$ Hz or $f=10^4$ Hz. The current varied from $10^{-3}$ to $5\times 10^{-2}$ A. An induction coil in Fig. 1(a) with the same parameters was wound on the other end of the LFC (detecting coil, $D$) to serve as a magnetic detector. The ac voltage in the coil was registered with a frequency-selective voltmeter. Figure 2(b) shows a system of two similar (up to 1.5 mm long) coils positioned apart at the distance equal to the LFC length (with an air gap between them) that was used for reference.

The voltage $V$ of the detector was measured in the course of the experiment as a function of the LFC length ($L$). The length varied from several millimeters to 90 mm. The dependence $V(L)$ was measured on the LFC of different diameters at different magnetic moments $M_G$ of the generating coil ($M_G$ is proportional to the current $I$ through the coil). The LFC diameter and the frequency of the generated field were chosen so that the LFC radius could be smaller than the skin depth at these parameters. The LFC-induced amplification $A$ of the dipole field was calculated using a simple relation $A=V/V0$ where $V0$ is the voltage of the detector in the reference system (an air gap between the coils in the absence of LFC).

## III. RESULTS AND DISCUSSION

The typical dependences $V(L)$ and $V0(L)$ for the field-exciting coil with $I=10^{-3}$ and $5\times 10^{-2}$ A are shown in Fig. 2. Similar curves were obtained for current region between $10^{-3}$ and $5\times 10^{-2}$ A, which corresponds to the magnetic moments of the coil $M_G=5\times 10^{-6}-2.5\times 10^{-4}$ A m$^2$. These dependences were used to calculate the amplification $A$ with different LFCs. The dependences $A(L)$ for the minimum and maximum magnetic moments $M_G$ of the coil at different LFC diameters and two frequencies $f=10^3$ Hz and $f=10^4$ Hz are illustrated in Fig. 3. The typical features of

[a)]Author to whom correspondence should be addressed; FAX: 38-057-3410933; electronic mail: bondarenko@ilt.kharkov.ua

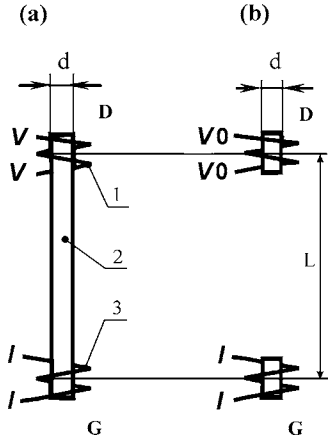

FIG. 1. Schematical views of (a) device with a concentrator for the experiments (1, generating coil $G$ for simulation of the influence of local magnetic field on a concentrator; 2, concentrator; 3, detecting coil $D$); (b) reference device without a concentrator.

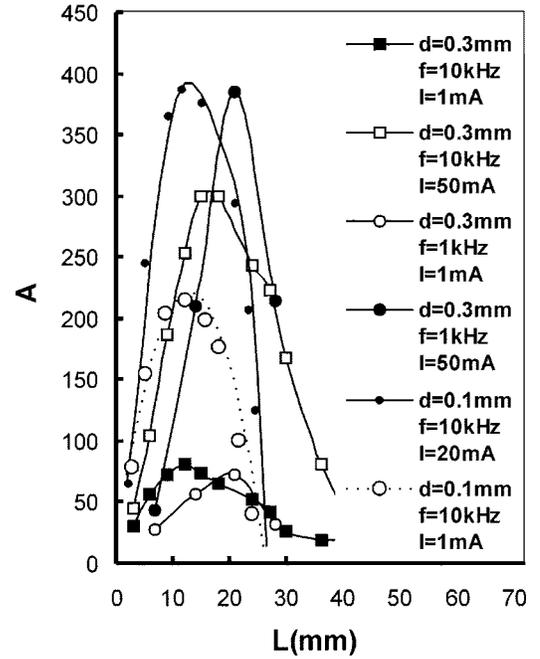

FIG. 3. Dependences of concentration $A$ of the magnetic field of the generating coil upon the concentrator length $L$ at different coil currents, concentrator diameters (0.1 and 0.3 mm), and current frequencies (1 and 10 kHz).

these dependences are the growth of $A$ in the region of a small LFC length, the maximum amplification $A_{max}$ at $L=12-20$ mm, and the subsequent decrease in $A$ at a large LFC length. $A_{max}$ increases with $M_G$ but this growth ceases at $M_G \approx 2.5 \times 10^{-4}$ A m$^2$ (Fig. 4). However, for one LFC diameter and one frequency of the current through the coil, a change in $M_G$ only slightly affects the position of the amplification maximum on the axis of the LFC lengths.

Finally, Fig. 5 demonstrates the dependences of the voltage in the detecting coil upon the magnetic moment $M_G$ of the LFC at the frequency $f=10^4$ Hz for two LFC diameters and three LFC lengths. It is seen that the dependence $V(M_G)$ remains linear in the whole range of $M_G$ values for the LFC with $d=0.3$ mm. This shape of the curve is also observed for three times smaller $M_G$ if the LFC diameter is one-third as much ($d=0.1$ mm).

It is interesting to consider the processes in the LFC exposed to the local field of the generating coil. The voltage $V$ of the detecting coil is proportional to the magnetic flux $\Phi$ through this part of the LFC. Since the LFC diameter is small, we can assume that the magnetic induction $B$ of the dipole field at this point of the LFC is about $B \approx \Phi/S$, where $S$ is the LFC cross-section area. Correspondingly, if there is no LFC between the dipole and the detector, the voltage $V0$ is proportional to the induction $B_0$ at the same point stimulated by a dipole having the same magnetic moment $M_G$. Since $B_0=\mu_0 H$, $B=\mu_0 \mu H$, where $\mu_0=4\pi 10^{-7}$ H/m, $\mu$ is the effective permeability of LFC, $H$ is the magnetic field value, the LFC-induced amplification of the dipole field is

$$A = B/B_0 = V/V0 = \mu. \tag{1}$$

The analysis of experimental data shows that the interpretation of the dependences $A(L)$, $A_{max}(M_G)$, and $V(M_G)$ as well

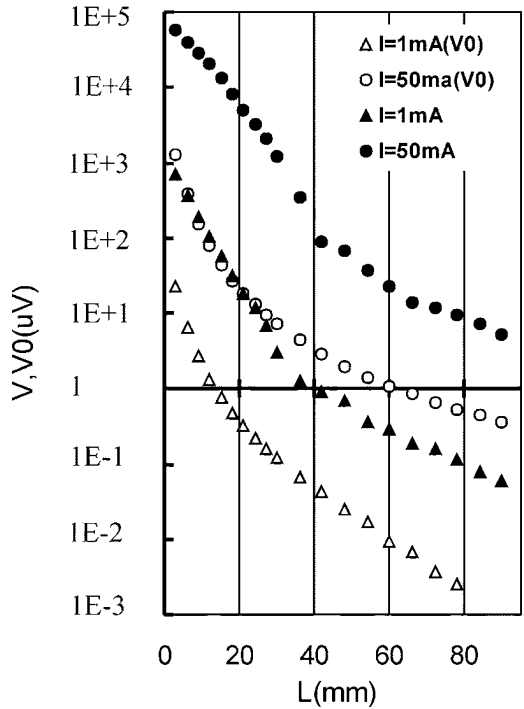

FIG. 2. Voltage $V$ of the detecting coil ($D$) vs the concentrator length $L$ [see Fig. 1(a)] and voltage $V0$ vs the distance $L$ for the configuration in Fig. 1(b) at different currents $I$ through the generating coil ($G$); the concentrator diameter ($d$) is 0.3 mm and the frequency of the current is 10 kHz.

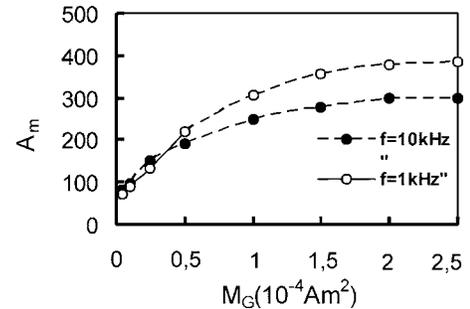

FIG. 4. Dependence of the maximum concentration $A_m$ of the generating coil field upon magnetic moment $M_G$ of the coil $G$ at two coil current frequencies (1 and 10 kHz) for the concentrator with diameter of 0.3 mm.

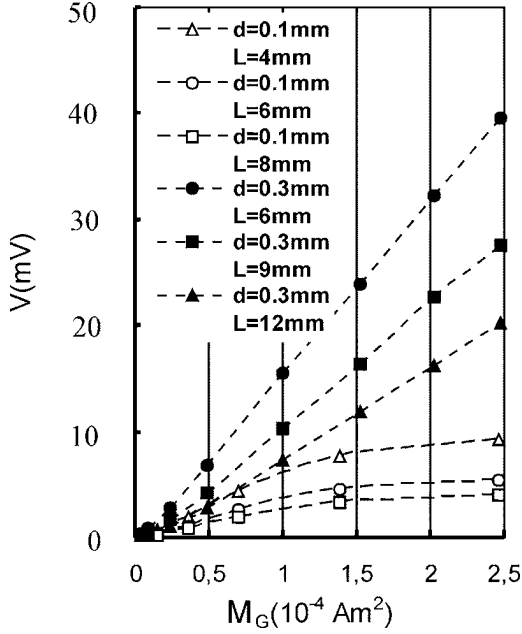

FIG. 5. Dependences of the detecting coil voltage $V$ upon the magnetic moment $M_G$ of the generating coil $G$ at different lengths $L$ and concentrator diameters $d$ (the current frequency in the generating coil is 10 kHz).

as the calculation of $A$ can be carried out using the known[6] relation

$$\mu \approx \mu_m/(1 + N\mu_m), \qquad (2)$$

supplemented with certain refinement and assumption concerning $\mu_m$ and $N$. First, Eq. (2) defines $\mu_m(H)$ as permeability of the LFC material corresponding to $H$ at the point $L_0$ separated from the dipole by the distance equal to one-half of the LFC length ($L_0 = 0.5L$). Second, according to Eq. (2), the demagnetization coefficient $N$ of LFC in an inhomogeneous field coincides with that for LFC in a homogeneous field. The magnitude $N$ is determined by the relation[6]

$$N = [\ln(1.2L/d) - 1]/(L/d)^2, \qquad (3)$$

if $L/d > 1$.

At the beginning we check that Eq. (2) is valid for a qualitative explanation of the experimental dependences. For short ($L < 12$ mm) LFC, when the coefficient $N$ is high and the relation $N \gg 1/\mu_m(H)$ is obeyed, amplification is only determined by the dimensions $L/d$ of LFC, being $A = \mu = 1/N$. When the length of such LFC increases from 3 to 12 mm, $A$ starts to grow (Fig. 3) in response to decreasing $N$. The intensity of the dipole field at the point $L_0$ is much higher than the coercive force of the LFC material. Correspondingly $\mu_m(H)$ is much lower than the maximum permeability $\mu_{m,max}$ of the LFC material. With LFC length of 12–20 mm, the $N$ value decreases and becomes commensurable with $1/\mu_m(H)$ at the point $L_0$. The magnitude $[\mu = A$ reaches its maximum ($A_m$) when the equation $1/\mu_m(H) = N$ is obeyed. These lengths correspond to a certain value $L = L_{max}$. Figure 3 shows, in particular, that $L_{max} = 20$ mm at $f = 10^3$ Hz, $d = 0.3$ mm. The highest $\mu_m(H)$ is attainable in a field close to the coercive force of the LFC material. As follows from Fig. 4, $A_m$ can amount to 70% of $\mu_{m,max}$ of the LFC material. On a further increase in the LFC length, the value of the dipole field at the point $L_0$ becomes smaller than the coercive force of the LFC material, and another inequality comes into play: $1/\mu_m(H) \gg N$. As a result, the amplification $A$ decreases. Equation (2) also explains how $A(L)$ is influenced by the changes in the LFC diameter and the dipole field frequency. For LFC of a smaller diameter (0.1 mm), the amplification maximum is expected to shift on the length axis towards smaller lengths because the quantities $N$ and $L/d$ required to meet the equation $1/\mu_m = N$ are attainable at a smaller value of $L_{max}$. Precisely such shift was observed experimentally (Fig. 3). A decrease in the frequency of the dipole field (from $10^4$ to $10^3$ Hz) usually leads to higher permeability $\mu_m$ in soft ferromagnetic material. The same is true for the LFC material. Hence, the equation $1/\mu_m = N$ is valid at smaller $N$ (with a larger ratio $L/d$). In this case the maximum in the curve $A(L)$ should shift towards a larger length $L_{max}$. This shift is obvious in Fig. 3. Finally, as seen in Fig. 5, for small-length LFC ($L/d<40$) the voltage $V$ of the detector of LFC lengths corresponds to the LFC permeability when $1/\mu_m \ll N$. This type of the dependence $V(M_G)$ is most suitable when we consider the application of LFC in magnetic microscopes for precise identification of sources generating local fields at the surface of the object in question.

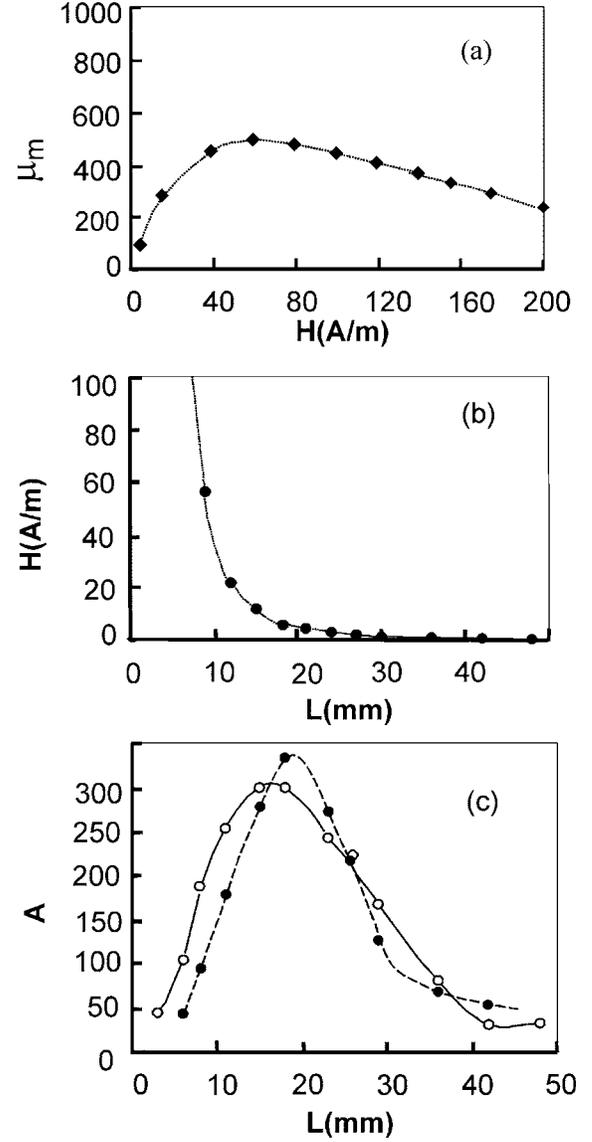

FIG. 6. (a) The experimental dependence of the magnetic permeability $\mu_m$ of the concentrator material (Permalloy wire) upon the magnetic field $H$ of 10 kHz. (b) Calculated distribution of the magnetic field $H$ of the generating coil with current $I = 50$ mA at $f = 10$ kHz along the concentrator (0.3 mm in diameter) length $L$. (c) Experimental (°) and calculated (●) dependences of concentration $A$ of the field of the generating coil with $I = 50$ mA, $f = 10$ kHz upon the 0.3 mm diameter concentrator length $L$.

Thus, the above consideration accounts quite reasonably for the main features of experimental curves.

The quantitative correlation between the experimental results and the calculation of $A(L)$ by Eq. (2) was analyzed using the following parameters: $f=10^4$ Hz, $M_G=2.5\times 10^{-4}$ A m$^2$ ($I=50$ mA). The value of $\mu_m(H)$ of the LFC wire material was measured in a homogeneous ac field with the frequency $f=10^4$ Hz using the induction technique [Fig. 6(a)]. The value of the dipole field was estimated using the relation $H(L)=M_G/2\pi L^2$ [Fig. 6(b)]. Figure 6(c) illustrates the experimental (○) and calculated (•) dependences $A(L)$. The curves are in approximate agreement.

Thus, our investigation of LFC prepared from soft magnetic material has shown that the value of amplification at frequencies of $10^3$ and $10^4$ Hz of the strongly inhomogeneous magnetic field at the LFC end can be explained and calculated using Eq. (2), which postulates that the demagnetization coefficient of LFC is identical with that in a homogeneous field. The value of permeability $\mu_m(H)$ of the LFC material correlates with the value of the dipole field ($H$) at the point corresponding to half-length of LFC. It is reasonable to expect that the conclusions drawn are valid at lower frequencies of the dipole field.

On the basis of the obtained outcomes we have manufactured concentrators with different parameters ($d=0.1$ mm, $L=3-5$ mm). The experiments carried out by us have shown that the sensitivity of SQUID detectors to a local permanent field can be increased by these concentrators in five to ten times.

### ACKNOWLEDGMENTS

The authors wish to thank N. Scherbakova and D. Fomin for technical support.